\pdfoutput=1
\documentclass[reprint,12pt,onecolumn,notitlepage,nofootinbib,floatfix]{revtex4-2}
\usepackage{amssymb}
\usepackage{amsmath}
\usepackage{amsfonts} 
\usepackage{bm}
\usepackage{color} 
\usepackage{subfigure}
\usepackage{graphicx}
\usepackage[bookmarks=false]{hyperref} 
\hypersetup{pdfstartview=FitH,pdfhighlight=/O,colorlinks=true}

\allowdisplaybreaks

\usepackage{epsfig} 
\usepackage{epstopdf}
\usepackage{latexsym}
\usepackage{psfrag}   
\usepackage{subfigure}  
\usepackage{booktabs}  
\usepackage{braket}
\usepackage{mathtools}
\usepackage{textcomp}
\usepackage{ifthen}
\usepackage{tensor}
\usepackage{protosem} 
\usepackage{wasysym}

\usepackage[toc]{appendix}
\usepackage{color,soul} 
\usepackage{datetime}

\definecolor{darkblue}{rgb}{0.2, 0, 0.8}
\definecolor{darkgreen}{rgb}{0.2, 0.71, 0}
\definecolor{awesome}{rgb}{1.0, 0.13, 0.32}
\definecolor{cadmiumred}{rgb}{0.89, 0.0, 0.13}
\definecolor{dukeblue}{rgb}{0.0, 0.0, 0.61}

\newcommand{\bea}{\begin{eqnarray}}
\newcommand{\eea}{\end{eqnarray}}
\newcommand{\ba}{\begin{eqnarray}}
\newcommand{\ea}{\end{eqnarray}}

\newcommand{\beq}{\begin{equation}}
\newcommand{\eeq}{\end{equation} }
\newcommand{\beqa}{\begin{eqnarray}}
\newcommand{\eeqa}{\end{eqnarray}}
\newcommand{\beqar}{\begin{eqnarray*}}
\newcommand{\eeqar}{\end{eqnarray*}}

\renewcommand{\href}[2]{#2}

\begin{document}

\title{Stochastic origin of primordial fluctuations in the Sky}

\author{Sayantan Choudhury}
\email{sayanphysicsisi@gmail.com, schoudhury@fuw.edu.pl, sayantan.choudhury@nanograv.org (Corresponding Author)}
\affiliation{Institute of Theoretical Physics, Faculty of Physics,
University of Warsaw, ul. Pasteura 5, 02-093 Warsaw, Poland }

\date{\today}

\begin{abstract}  \vskip 0.2in 

We provide a study of the effects of the Effective Field Theory (EFT) generalisation of stochastic inflation on the production of primordial black holes (PBHs) in a model-independent single-field context. We demonstrate how the scalar perturbations' Infra-Red (IR) contributions and the emerging Fokker-Planck equation driving the probability distribution characterise the Langevin equations for the ``soft" modes in the quasi-de Sitter background. Both the classical-drift and quantum-diffusion-dominated regimes undergo a specific analysis of the distribution function using the stochastic-$\delta N$ formalism, which helps us to evade a {\it no-go theorem} on the PBH mass. Using the EFT-induced alterations, we evaluate the local non-Gaussian parameters in the drift-dominated limit.

\vskip 0.2in
\centering
\noindent {\it \footnotesize Essay written and received honorable mention for the Gravity Research Foundation 2025 Awards for Essays on Gravitation}

\end{abstract}   

\maketitle 

\newpage

\section{Introduction}

One of the most prominent models for the very early universe is cosmological inflation, which offers a seeding process for the creation of large-scale structures of today from primordial quantum fluctuations. These fluctuations, which are often linked to a scalar field that drives inflation, undergo a transition from the quantum domain to the large-scale, classical domain. The dynamics of large-scale fluctuations that are impacted in the presence of noise components originating from the quantum-to-classical transition of the tiny wavelength modes of primordial fluctuations were previously studied using the stochastic inflationary paradigm. Primordial black hole (PBH) production \cite{Zeldovich:1967lct,Hawking:1974rv,Carr:1974nx,Carr:1975qj,Vennin:2020kng,Riotto:2023hoz,Riotto:2023gpm,Papanikolaou:2022did,Choudhury:2011jt,Choudhury:2023vuj, Choudhury:2023jlt, Choudhury:2023rks,Choudhury:2023hvf,Choudhury:2023kdb,Choudhury:2023hfm,Bhattacharya:2023ysp,Choudhury:2023fwk,Choudhury:2023fjs,Choudhury:2024one,Firouzjahi:2023ahg,Firouzjahi:2023aum,Riotto:2024ayo,Choudhury:2024dei,Choudhury:2024dzw,Choudhury:2024aji,Choudhury:2024kjj,Choudhury:2024ybk} is an intriguing byproduct of the primordial oscillations in the early universe, namely those produced at tiny scales close to the end of inflation. PBHs are produced by the gravitational collapse of regions of overdensities and underdensities in the universe's substance that result from significant fluctuations at smaller scales and soon after they re-enter the horizon. Among the potential processes for PBH production is the development of a nearly flat zone close to the smaller field values of the inflationary potential, which greatly amplifies the field fluctuations taking part formation of PBHs. This inflationary regime is now commonly referred to as an ultra-slow roll (USR) phase, following horizon escape, when quantum diffusion effects become equally significant and contribute to the overall dynamics of the large-scale classical perturbations.

Through the development of a soft de Sitter Effective Field Theory (SdSET) \cite{Gorbenko:2019rza,Cohen:2021fzf,Cohen:2022clv,Green:2022ovz,Cohen:2021jbo,Cohen:2020php} formulation of stochastic single-field inflation \cite{Starobinsky:1986fx,Vennin:2024yzl,Animali:2024jiz,LISACosmologyWorkingGroup:2023njw,Animali:2022otk,Ezquiaga:2022qpw,Jackson:2022unc,Tada:2021zzj,Pattison:2021oen,Ando:2020fjm,Vennin:2020kng,Ezquiaga:2019ftu,Pattison:2019hef,Noorbala:2018zlv,Pattison:2017mbe,Grain:2017dqa,Hardwick:2017fjo,Mishra:2023lhe,Choudhury:2024jlz}, we hope to generalise this image in a model-independent manner without explicitly introducing any scalar fields into the framework. When the gauge-invariant variable, or the comoving curvature perturbation in the EFT of inflation \cite{Weinberg:2008hq,Cheung:2007st,Choudhury:2017glj,Delacretaz:2016nhw,Naskar:2017ekm}, is separated into its long and short wavelength components, the low-energy component is referred to as ``soft" in this context. Stochastic influences would cause the short-wavelength components to coarsen, and following horizon crossing, they would subsequently join the long-wavelength dynamics. When stochastic influences are present, the Ultra-Violet (UV) component of the fluctuations experiences a process of coarse-graining during the instant of horizon crossing. Additionally, depending on the coarse-graining window function selected, this instant is not precise. The horizon crossing is shown by a correct moment in the time coordinate when a Heaviside Theta function is used. The use of this function produces what are known as ``white" noise terms, which give the system a Markovian description. There are sounds which are labelled coloured noises, and is indicative of a non-Markovian system, when a window function is specified using a certain profile. The stochastic Langevin equation describes the evolution of the coarse-grained curvature perturbations, commonly known as the Infra-Red (IR) component, and this crossover process stretches into the super-Hubble scales. 

A perturbative technique cannot be expected to provide adequate knowledge of PBH creation as it becomes inevitable to look beyond the assumed Gaussian statistics for the curvature perturbation. As a result, in order for the huge perturbations to have the greatest impact, the distribution must exhibit notable tail characteristics and departures from Gaussianity. In the USR, we use this strategy and demonstrate how the Fokker-Planck equation for the expansion variable's probability distribution function (PDF) works. This work uses the stochastic-$\delta N$ formalism \cite{Enqvist:2008kt,Fujita:2013cna,Fujita:2014tja,Vennin:2015hra} to examine stochastic effects in a USR environment.

\section{Stochastic EFT for PBH formation}\label{sec2}

The process of incorporating stochasticity into the EFT formalism entails first analysing the stochastic nature of fluctuations using Hamilton's equations, or more specifically, the Langevin equations \cite{Grain:2017dqa,Vennin:2020kng}. Next, using the Fokker-Planck equation, the distribution function of the curvature perturbations is evolved from the Langevin equations.

\begin{figure*}[htb!]
    	\centering
    {
       \includegraphics[width=14cm,height=7.5cm]{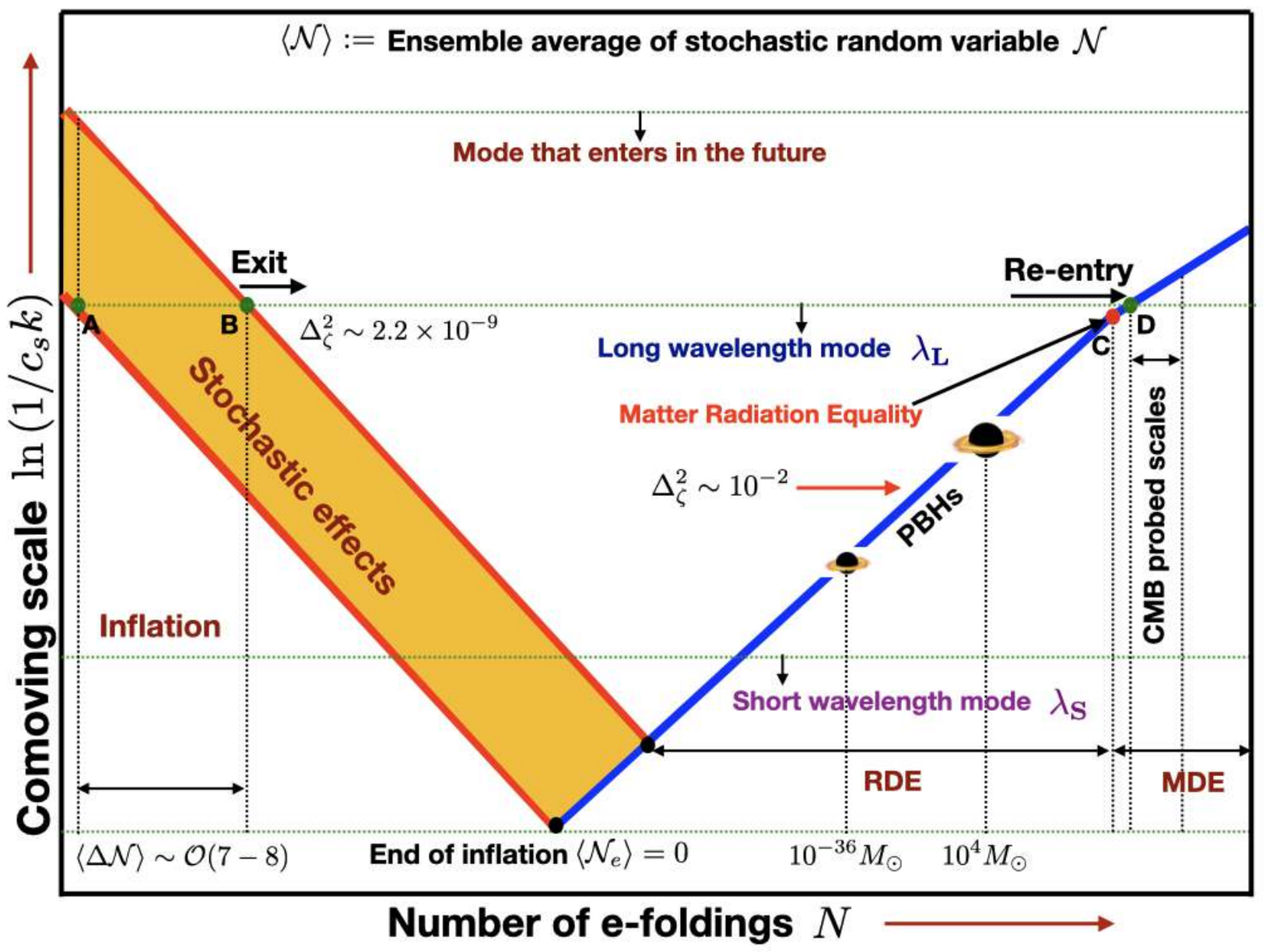}
    } 
    \caption[Optional caption for list of figures]{This diagram shows that during inflation, modes go from the Sub-Horizon to the Horizon crossing, where they face stochastic effects before re-entering the Horizon.   }
\label{stochastic}
    \end{figure*}
In the classical picture, Hamilton's equations of motion in our EFT picture are described by the following equations:
\bea \label{Langevin1}
\frac{d\zeta}{dN} &=& \Pi_{\zeta}, \quad\quad
\label{Langevin2}
\frac{d\Pi_{\zeta}}{dN} =  -(3-\epsilon)\Bigg[1 - \frac{2(s-\frac{\eta }{2})}{(3-\epsilon)} \Bigg]\Pi_{\zeta},
\eea
where $\zeta$ and $\Pi_{\zeta}$ describe the scalar curvature perturbation and its associated canonically conjugate momenta. Other symbols describe slow-roll parameters, which are defined as, $s = \frac{d\ln c_{s}}{dN},\epsilon = -\frac{d\ln{H}}{dN},\eta = \epsilon-\frac{1}{2}\frac{d\ln{\epsilon}}{dN}$. Here, $N$ describes the number of e-folds. 

Now, moving to the quantum picture, it is useful to separate the curvature perturbations into the two distinct UV and IR components, to derive the Langevin equations for the coarse-grained components of the initial quantum fields, which is written as, $\Hat{\Gamma} = \Hat{\Gamma}_{\bf IR} + \Hat{\Gamma}_{\bf UV}$ where $\Hat{\Gamma}_{\bf IR}= \{\Hat{\zeta},\Hat{\Pi}_{\zeta}\}, \Hat{\Gamma}_{\bf UV}= \{\Hat{\zeta}_{s},\Hat{\Pi}_{\zeta_{s}}\},$. Here, the subscript $s$ corresponds to the small-wavelength contribution. In this context, the UV mode can be expressed in terms of Fourier modes:
\bea \Hat{\Gamma}_{\bf UV} &=& \int_{\mathbb{R}^3}\frac{d^3 k }{(2\pi)^3 }W\Bigg(\frac{k}{k_\sigma}\Bigg)\Bigg[ \Hat{a}_{{\bf k}}\Hat{\Gamma}_{k}(\tau) e^{-i {\bf k}.{\bf x}} +h.c.\Bigg]\quad{\rm where}\quad \Hat{\Gamma}_{k}(\tau)=\{\zeta_{k}(\tau),\Pi_{\zeta_k}(\tau)\}.\eea
Here $k_{\sigma}=\sigma aH$, where $\sigma$ is the coarse-graning parameter that controls the stochasticity and the window function $W$ we choose as the Heaviside Theta function. Also, $\Hat{a}_{{\bf k}}$ and its conjugate satisfy the standard canonical commutation relations. The Langevin equations for the UV modes in the quantum picture are described as:
\bea 
\frac{d\hat{\zeta}}{dN} &=&  \hat{\Pi}_{\zeta} + 
 \hat{\xi}_\zeta(N),\quad\quad 
\frac{d\hat{\Pi}_\zeta }{dN} = -(3-\epsilon)\hat{\Pi}_\zeta \Bigg[1 -\frac{2(s-\frac{\eta }{2})}{(3-\epsilon)} \Bigg] + \hat{\xi}_{\pi_\zeta}(N),
\eea 
The quantum white noise terms, which are supplied due to the continuous escape of UV modes into the IR regime, are indicated using $\hat{\xi}_\zeta(N)$ and $\hat{\xi}_{\pi_\zeta}(N)$ and are described in terms of the Fourier modes as:
\bea 
\Hat{\xi}_{\Gamma} &=& -\int_{\mathbb{R}^3}\frac{d^3 k }{(2\pi)^3 }\frac{d}{dN}W\Bigg(\frac{k}{k_\sigma}\Bigg)\Bigg[ \Hat{a}_{{\bf k}}\Hat{\Gamma}_{k}(\tau) e^{-i {\bf k}.{\bf x}} + h.c.\Bigg]\quad{\rm where}\quad \Hat{\xi}_{\Gamma}=\{\hat{\xi}_\zeta(N),\hat{\xi}_{\pi_\zeta}(N)\}.\eea
   
In figure (\ref{stochastic}), we have shown that corresponding to the CMB scales, the large-scale $\lambda_{L}$ modes leave at positions $A$ and $B$ and re-enter at the moment $D$. The CMB re-entry scale is near the radiation-matter equality, denoted using the symbol $C$. The shorter length scales towards the end of inflation, which contribute to the subsequent collapse to form PBHs during radiation domination with varying masses, are linked to the short wavelength $\lambda_{S}$. The number of e-folds becomes a stochastic variable ${\cal N}$ that passes between a beginning and final set of circumstances.

\section{A short note on stochastic $\delta N$ formalism in EFT}
\begin{figure*}[htb!]
    	\centering
    {
       \includegraphics[width=14cm,height=7.2cm]{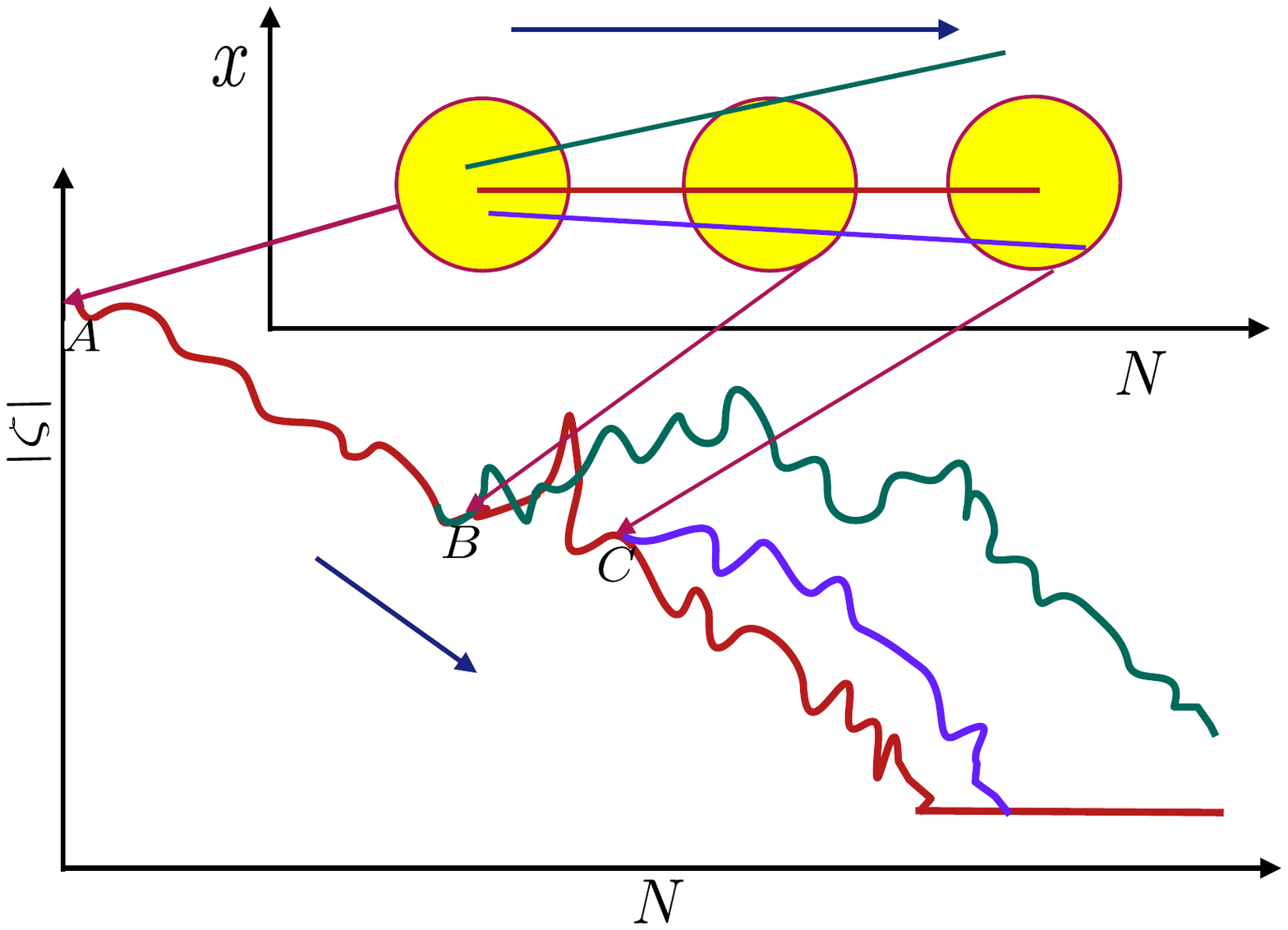}
    } 
    \caption[Optional caption for list of figures]{A schematic changing behaviour of coarse-grained curvature perturbation with e-folds $N$.  }
\label{deltaNstoc}
    \end{figure*}
Here, we do not explore a single FLRW Universe, but rather a family of them that evolve with the phase-space variables following a set of beginning conditions. An example of a phase space vector that combines these variables is ${\Gamma}_{i}=\{\zeta_{i},\Pi_{\zeta,i}\}$, where the index $i$ labels the distinct components. A low-energy EFT with an IR component of the first primordial fluctuations is constructed as part of the stochastic formalism using the curvature perturbations. The coarse-grained versions of these IR modes are denoted by $\zeta_{\rm cg}({\bf x}) = \int_{\mathbb{R}^3, k<k_{\sigma}}\frac{d^3 k }{(2\pi)^3 }\zeta_{k}e^{-i{\bf k}.{\bf x}}$. A large amount of short-scale modes engage in the zone of stochastic effects, become ``classicalized," and eventually join the IR sector as the horizon size keeps decreasing. The consequence is the birth of classical noises, which in turn regulate the super-Hubble modes' evolution, which are explained by the Langevin equation, the results of which will be examined in the next section in terms of PDFs.

The quantity ${\cal N}$, which corresponds to the total expansion achieved along the worldline trajectory for a point—from an initial condition to some final hypersurface—behaves as a stochastic variable. Using the stochastic-$\delta N$ formalism, the coarse-grained version of the IR modes can be represented as, $\zeta_{\rm cg}({\bf x}) = {\cal N}({\bf x})-\langle{\cal N}\rangle = \delta {\cal N}.$ Consequently, following the solution of the Langevin equation for several realisations at a certain spatial position, the angle brackets then indicate a statistical average. In figure (\ref{deltaNstoc}), the Gaussian random noises begin to impact points B and C individually, and the development of spatial points during inflation is further depicted in the same figure.
At every spatial location inside the Hubble patch (yellow circles), the coarse-grained curvature perturbation remains the same, beginning at the point $A$ and continuing up to $B$ and $C$. Following their exit, they develop statistically independently (green and blue lines).

Using the stochastic $\delta N$ formalism, in terms of the various statistical moments $\langle\delta{\cal N}^{q}\rangle = \langle({\cal N}-\langle{\cal N}\rangle)^q\rangle$, where $q=2,3,4$, the dimensionless power spectrum and other non-Gaussian amplitudes are estimated by the following relations:
\bea  &&\Delta^{2}_{\delta N} = \frac{d}{d\langle{\cal N}\rangle }\langle\delta{\cal N}^{2}\rangle \big|_{\langle{\cal N}\rangle=\ln(k_{f}/k)}, f_{\rm NL} = \frac{5}{36}\frac{d\langle\delta{\cal N}^{3}\rangle^2}{d\langle{\cal N}\rangle^2}\bigg(\frac{d\langle\delta{\cal N}^{2}\rangle}{d\langle{\cal N}\rangle}\bigg)^{-2},\nonumber\\
&&\tau_{\rm NL} =\frac{1}{36}\bigg(\frac{d^2\langle\delta{\cal N}^3\rangle}{d\langle{\cal N}\rangle^2}\bigg)^{2}\bigg(\frac{d\langle\delta{\cal N}^{2}\rangle}{d\langle{\cal N}\rangle}\bigg)^{-4},\quad
g_{\rm NL} = \frac{d\langle\delta{\cal N}^{4}\rangle^3}{d\langle{\cal N}\rangle^3}\bigg(\frac{d\langle\delta{\cal N}^{2}\rangle}{d\langle{\cal N}\rangle}\bigg)^{-3}.\eea

\section{PDFs and PBHs as an outcome of Fokker-Planck equation}

The Fokker-Planck equation in the present EFT setup can be expressed as:
\bea \label{FPE}
\frac{\partial }{\partial \mathcal{N} }P_{\Gamma_i}(\mathcal{N})&=& \bigg[\frac{1}{\tilde{\mu}^{2}}\frac{\partial^{2}}{\partial x^{2}} - 3Cy\bigg\{\frac{\partial}{\partial x} + \frac{\partial}{\partial y}\bigg\} \bigg]P_{\Gamma_i}({\cal N}), P_{{\Gamma_i}}({\cal N}) = \frac{1}{2\pi}\int_{-\infty}^{+\infty}e^{-it{\cal N}}\chi(t;{\Gamma_i})dt.\eea
Here $P_{{\Gamma_i}}({\cal N})$ represents PDFs which are related to the characteristic function $\chi(t;{\Gamma_i})= \sum_{n=0}^{\infty}\frac{(it)^{n}}{n!}\langle{\cal N}^{n}({\Gamma_i})\rangle$. We also define, ${\Gamma_i} = \{\zeta,\Pi_{\zeta}\}$, $\tilde{\mu}=C/\mu$, $x=C\zeta$, $y=-\Pi_{\zeta}/3$ and the EFT-based parameter $C=\left(1-\frac{\epsilon}{3}\right)\left(1-\frac{2(s-\frac{\eta}{2})}{3-\epsilon}\right)$.

The desired solution of the Fokker-Planck equation in terms of PBH formation can be found in the following two regimes:
\begin{enumerate}
    \item \underline{\bf  Quantum diffusion-dominated regime:}
    In this region, the dynamics are controlled by the quantum effects. In particular, the canonically conjugate momentum variable lies within, $0<y\ll 1$. Because we will be witnessing distinguishing features in the upper tail of the PDF—the region most pertinent for the generation of PBH—this interval is important from the viewpoint of a detailed PBH analysis. Moreover, it can be understood using the conjugate field momenta as it diminishes in this regime; hence, the scalar perturbations and PDF characteristics are predominantly governed by diffusion effects. The desired analytical form of the PDF in this region is illustrated by the following equation:
\bea 
P_{\Gamma_i}({\cal N}) = -i\sum_{m=0,1,\cdots}\sum_{n=0}^{\infty}\bigg[\left.\frac{\partial}{\partial t}\chi^{-1}(t;{\Gamma_i})\right\vert_{t=-i\Lambda_{n}^{(m)}}\bigg]^{-1}\exp{(-\Lambda_{n}^{(m)}{\cal N})},
\eea
where $\Lambda_{n}^{(m)} = 3Cm + \left[\frac{\pi}{\tilde{\mu}}\left(n+\frac{1}{2}\right)\right]^{2}$.
    
    \item \underline{\bf Classical drift-dominated regime:}
Here, we focus on the opposite scenario, in which drift effects for the scalar perturbations become crucial and their dynamics through USR is primarily controlled by the classical limit where quantum diffusion proves to be subdominant. For a given $\Gamma_{i}$, diffusion processes do not operate for the majority of e-folding realisations between some beginning and final conditions and hence do not primarily impact the scalar perturbations. The desired analytical form of the PDF in this region is represented by the following relation:
\bea \label{pdfLO1}
P_{\Gamma}({\cal N})= \delta\left({\cal N}+\frac{1}{3C}\ln{\left(1-\frac{x}{y}\right)}\right)+{\rm Non-Gaussian\quad contributions}.
\eea

\end{enumerate}
\begin{figure*}[ht!]
    	\centering
    \subfigure[]{
      	\includegraphics[width=7.9cm,height=6cm]{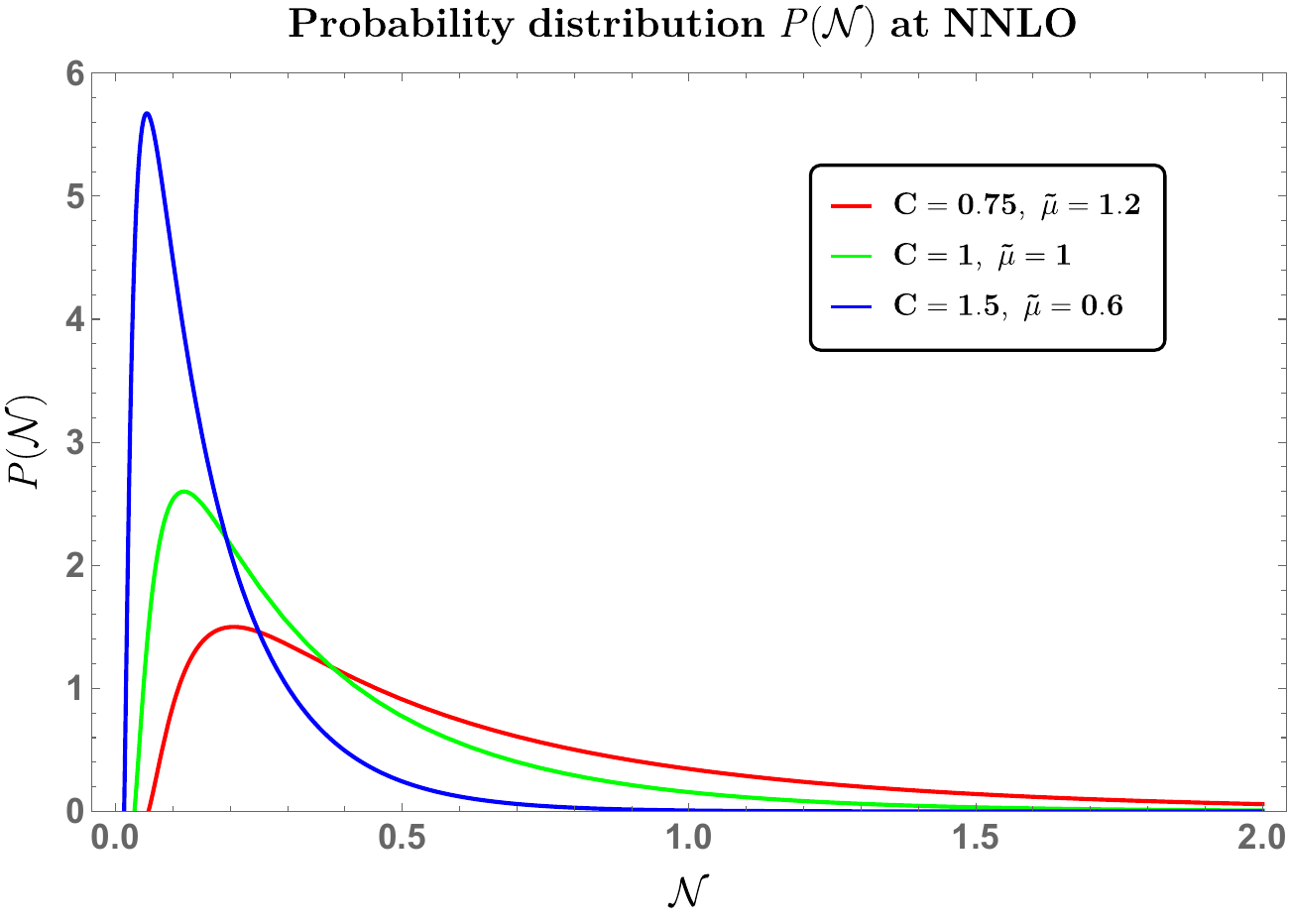}
        \label{f3a}
    }
    \subfigure[]{
        \includegraphics[width=7.9cm,height=6.2cm]{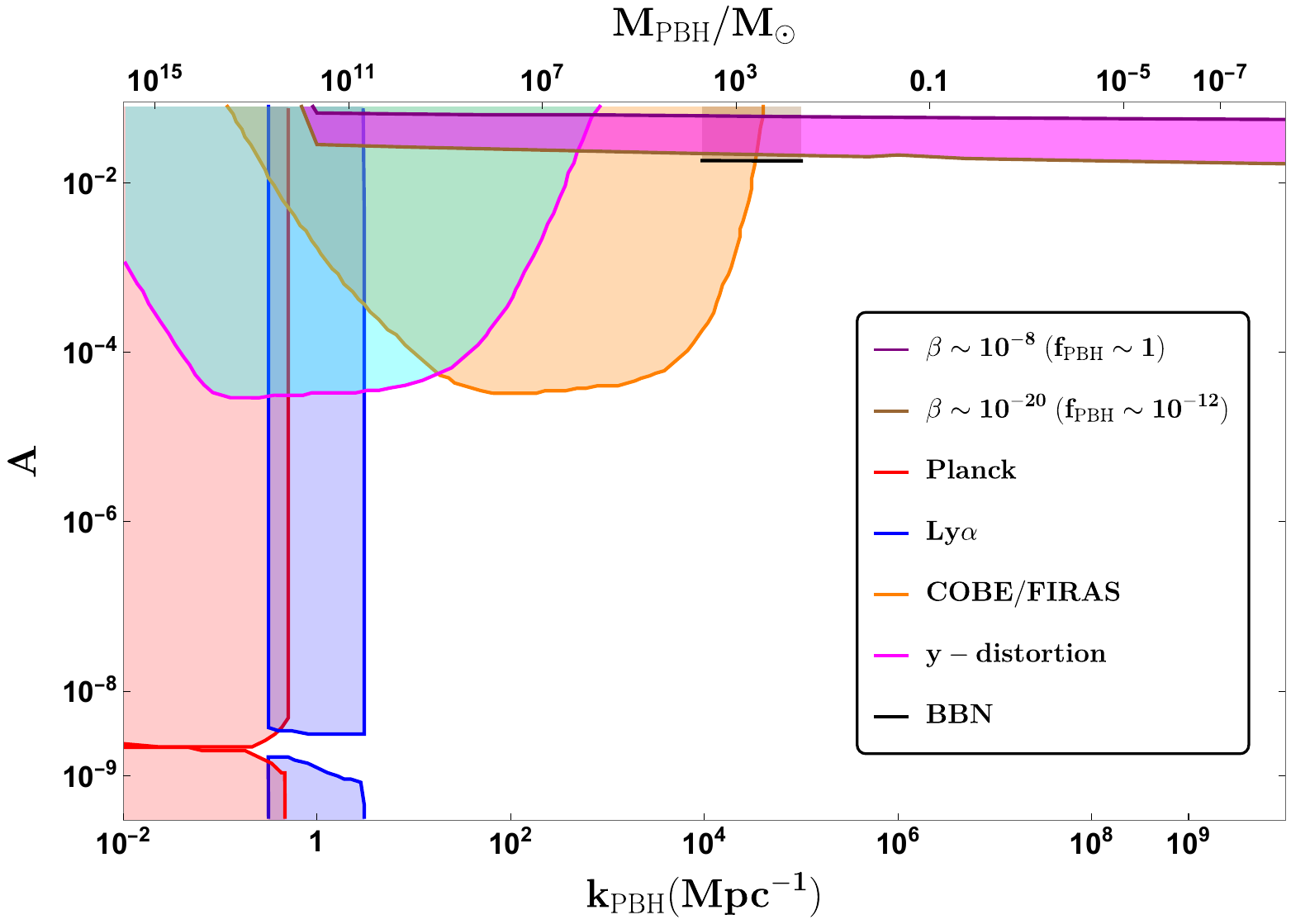}
        \label{f3b}
    }
    	\caption[Optional caption for list of figures]{Behaviour of the \ref{f3a} PDF with stochastic variable ${\cal N}$ and \ref{f3b} scalar power spectrum amplitude needed to attain a sufficient PBH mass fraction $\beta$ with wave number.} 
    	\label{dyn}
    \end{figure*}
Using this PDF, one can further compute the PBH mass fraction from $\beta \sim \int_{\zeta_{\rm th}+\langle{\cal N}\rangle}^{\infty}P_{\Gamma}({\cal N})d{\cal N}$, where the threshold is fixed at $\zeta_{\rm th}\sim {\cal O}(1)$. Here, this mass-fraction $\beta$ is also related to the PBH abundance $f_{\rm PBH}$, through the PBH mass $M_{\rm PBH}$ (i.e., $f_{\rm PBH}\propto M^{-1/2}_{\rm PBH}\beta$), which depends on the wave number $k_{\rm PBH}$ of PBH formation (i.e., $M_{\rm PBH}\propto k^{-2}_{\rm PBH}$).

In figure \ref{f3a}, we have shown the behaviour of the PDF with the stochastic variable ${\cal N}$ for the different values of the EFT-induced parameter $C$ and $\tilde{\mu}$. From the depicted features of PDFs, it is quite evident that non-Gaussian contributions are significant in the classical drift-dominated regime and can be estimated roughly as $f_{\rm NL}\sim 5C/2$, $\tau_{\rm NL}\sim C^2/4$, and $g_{\rm NL}\sim 1890 C^2$. The figure \ref{f3b} presents how the amplitude of the scalar power spectrum behaves when generated as an outcome of stochasticity and is necessarily required to produce the necessary and sufficient amount of PBH mass fraction $\beta$ for the corresponding wave number in which PBHs get produced. In this plot, we have also tagged the PBH masses produced for a better understanding purpose. In addition, constraints from various cosmological observations are present in different coloured shaded contour regions. In the allowed range, the required amplitude of the perturbation becomes, $A\sim {\cal O}(10^{-2})$, which allows $\beta\sim {\cal O}(10^{-20}-10^{-8})$, which corresponds to $f_{\rm PBH}\sim{\cal O}(10^{-12}-1)$. This allows us to generate PBH masses within the range $M_{\rm PBH}\sim {\cal O}(10^{-8}-10^{11})$ solar mass. For $M_{\rm PBH}>{\cal O}(10^{11})$ solar mass, the amplitude of the stochastic scalar perturbation increases beyond $A>0.1$, such that $f_{\rm PBH}>1$, which is completely disfavored in the present context of the analysis.

\section{ What is the big deal with the stochastic EFT framework?}

Moving on, an important question that comes to our mind next is that, earlier in many studies, even within the EFT framework, stochasticity has not been used to explain the generation of PBHs. So, what is the big deal with incorporating stochasticity in primordial fluctuations to explain the generation of PBHs? This is certainly a crucial question that is of concern in the present study, and we need to clarify it. In this section, we will try to give some of the important justifications regarding the usefulness of the stochasticity, which are appended pointwise below:
\begin{itemize}
    \item Incorporating stochasticity within the framework of EFT helps us to give a completely model-independent description to explain the generation of PBHs from large amplitude primordial fluctuations. It also helps to unify various single-field inflationary paradigms in a common description, which allows the inclusion of stochasticity to describe PBH formation. 

    \item Stochasticity is implemented with the help of the $\delta N$ formalism, which in itself describes a non-perturbative scenario. It has been previously noted that if we consider quantum loop effects and sum them up in all orders of perturbation theory with the help of the Dynamical Renormalization Group (DRG) resummation technique, this description exactly mimics the role of $\delta N$ formalism, thereby also describing a non-perturbative scenario \cite{Dias:2012qy}. Consequently, the stochastic $\delta N$ formalism within the framework of EFT gives the most accurate description of the formation of PBH from large primordial fluctuations.

    \item Due to having a specific type of interaction term appearing in the third-order interaction Hamiltonian of the cosmological perturbation theory, quantum loop effects along with DRG resummation lead to a {\it no-go theorem} on PBH mass \cite{Choudhury:2023vuj,Choudhury:2023jlt,Choudhury:2023rks,Choudhury:2024ybk,Choudhury:2024aji} which allows only to generate PBHs of mass $10^2$ gm and put stringent constraint on the single-field inflationary paradigm. However, the most promising outcome of the stochastic-EFT framework is that it bypasses such {\it no-go theorem} in the completely non-perturbative description and allows the generation of PBHs within a wide range having $M_{\rm PBH}\sim {\cal O}(10^{-8}-10^{11})$ solar mass. This result implies that the previously mentioned {\it no-go theorem} on PBHs mass can be evaded with the help of a stochastic-EFT scenario even though we have sharp or smooth slow-roll to ultra-slow-roll transitions and vice versa.

    \item With the help of the stochastic-$\delta N$ formalism \cite{Enqvist:2008kt,Fujita:2013cna,Fujita:2014tja,Vennin:2015hra} within the framework of EFT, non-Gaussian features have been very easily studied in the drift-dominated region of the obtained PDFs. On the other hand, the diffusion-dominated region describes the PBHs formation in terms of mass fraction $\beta$ from the tails of the PDFs \cite{Choudhury:2024jlz}. This is the strength and beauty of stochastic non-perturbative formalism, which helps us to determine two essential features of PDFs in a unified language under the umbrella of stochastic formalism, which is described in this essay. However, instead of using the stochastic non-perturbative formalism, if we use the usual cosmological perturbation theory in EFT setup, then determining the above-mentioned two crucial features from the amplitude of various correlation functions becomes extremely cumbersome and sometimes inconclusive. 

    \item Using stochastic formalism, one can further comment on the autocorrelation and cross-correlation amplitudes, which enter directly into the estimation of PDFs from the by-product of the Fokker-Planck equation. Most importantly, both correlations significantly contribute to the PDFs for estimating the non-Gaussianities and PBH formation. On the other hand, the perturbative calculations underestimates the cross-correlation functions, giving a negligible impact on the estimation of non-Gaussianities and PBH formation. As a result, in this particular context, stochastic non-perturbative formalism is more reliable than the results obtained from the usual perturbation theory performed in various orders.
\end{itemize}
\section{Discussion}
In this article, we present a model-independent analysis of the impact of the Effective Field Theory (EFT) extension of stochastic inflation on primordial black hole (PBH) generation within a single-field framework.
Using the stochastic-$\delta N$ formalism, we provide an elaborate justification of the usefulness of this non-perturbative scenario to study the non-Gaussian features in the drift-dominated regime and PBHs formation from the diffusion-dominated regime of the obtained PDFs utilizing the IR soft modes from the Fokker-Planck equation in a unified language. We have also justified in detail the utility of the stochastic-$\delta N$ formalism in the EFT framework over the usual perturbative computations. Most importantly, this approach represents one of the most successful proposals to evade the {\it no-go theorem}, which allows the generation of PBHs mass within the range $M_{\rm PBH}\sim {\cal O}(10^{-8}-10^{11})$ solar mass for the single-field inflationary paradigm. 

\emph{Acknowledgments:}
 SC sincerely thanks Ahaskar Karde for various useful discussions. SC acknowledges our debt to the people
belonging to various parts of the world for their generous and steady support for research in natural sciences.


\renewcommand{\leftmark}{\MakeUppercase{Bibliography}}
\phantomsection
\bibliography{Refs}

\providecommand{\href}[2]{#2}\begingroup\raggedright\begin{thebibliography}{10}

\bibitem{Zeldovich:1967lct}
Y.~B. Zel'dovich and I.~D. Novikov, ``{The Hypothesis of Cores Retarded during
  Expansion and the Hot Cosmological Model},'' {\em Soviet Astron. AJ (Engl.
  Transl. ),} {\bfseries 10} (1967) 602.

\bibitem{Hawking:1974rv}
S.~W. Hawking, ``{Black hole explosions},''
  \href{http://dx.doi.org/10.1038/248030a0}{{\em Nature} {\bfseries 248} (1974)
  30--31}.

\bibitem{Carr:1974nx}
B.~J. Carr and S.~W. Hawking, ``{Black holes in the early Universe},''
  \href{http://dx.doi.org/10.1093/mnras/168.2.399}{{\em Mon. Not. Roy. Astron.
  Soc.} {\bfseries 168} (1974) 399--415}.

\bibitem{Carr:1975qj}
B.~J. Carr, ``{The Primordial black hole mass spectrum},''
  \href{http://dx.doi.org/10.1086/153853}{{\em Astrophys. J.} {\bfseries 201}
  (1975) 1--19}.

\bibitem{Vennin:2020kng}
V.~Vennin, {\em {Stochastic inflation and primordial black holes}}.
\newblock PhD thesis, U. Paris-Saclay, 6, 2020.
\newblock \href{http://arxiv.org/abs/2009.08715}{{\ttfamily arXiv:2009.08715
  [astro-ph.CO]}}.

\bibitem{Riotto:2023hoz}
A.~Riotto, ``{The Primordial Black Hole Formation from Single-Field Inflation
  is Not Ruled Out},'' \href{http://arxiv.org/abs/2301.00599}{{\ttfamily
  arXiv:2301.00599 [astro-ph.CO]}}.

\bibitem{Riotto:2023gpm}
A.~Riotto, ``{The Primordial Black Hole Formation from Single-Field Inflation
  is Still Not Ruled Out},'' \href{http://arxiv.org/abs/2303.01727}{{\ttfamily
  arXiv:2303.01727 [astro-ph.CO]}}.

\bibitem{Papanikolaou:2022did}
T.~Papanikolaou, A.~Lymperis, S.~Lola, and E.~N. Saridakis, ``{Primordial black
  holes and gravitational waves from non-canonical inflation},''
  \href{http://dx.doi.org/10.1088/1475-7516/2023/03/003}{{\em JCAP} {\bfseries
  03} (2023) 003}, \href{http://arxiv.org/abs/2211.14900}{{\ttfamily
  arXiv:2211.14900 [astro-ph.CO]}}.

\bibitem{Choudhury:2011jt}
S.~Choudhury and S.~Pal, ``{Fourth level MSSM inflation from new flat
  directions},'' \href{http://dx.doi.org/10.1088/1475-7516/2012/04/018}{{\em
  JCAP} {\bfseries 04} (2012) 018},
  \href{http://arxiv.org/abs/1111.3441}{{\ttfamily arXiv:1111.3441 [hep-ph]}}.

\bibitem{Choudhury:2023vuj}
S.~Choudhury, M.~R. Gangopadhyay, and M.~Sami, ``{No-go for the formation of
  heavy mass Primordial Black Holes in Single Field Inflation},''
  \href{http://dx.doi.org/10.1140/epjc/s10052-024-13218-2}{{\em Eur. Phys. J.
  C} {\bfseries 84} no.~9, (2024) 884},
  \href{http://arxiv.org/abs/2301.10000}{{\ttfamily arXiv:2301.10000
  [astro-ph.CO]}}.

\bibitem{Choudhury:2023jlt}
S.~Choudhury, S.~Panda, and M.~Sami, ``{PBH formation in EFT of single field
  inflation with sharp transition},''
  \href{http://dx.doi.org/10.1016/j.physletb.2023.138123}{{\em Phys. Lett. B}
  {\bfseries 845} (2023) 138123},
  \href{http://arxiv.org/abs/2302.05655}{{\ttfamily arXiv:2302.05655
  [astro-ph.CO]}}.

\bibitem{Choudhury:2023rks}
S.~Choudhury, S.~Panda, and M.~Sami, ``{Quantum loop effects on the power
  spectrum and constraints on primordial black holes},''
  \href{http://dx.doi.org/10.1088/1475-7516/2023/11/066}{{\em JCAP} {\bfseries
  11} (2023) 066}, \href{http://arxiv.org/abs/2303.06066}{{\ttfamily
  arXiv:2303.06066 [astro-ph.CO]}}.

\bibitem{Choudhury:2023hvf}
S.~Choudhury, S.~Panda, and M.~Sami, ``{Galileon inflation evades the no-go for
  PBH formation in the single-field framework},''
  \href{http://dx.doi.org/10.1088/1475-7516/2023/08/078}{{\em JCAP} {\bfseries
  08} (2023) 078}, \href{http://arxiv.org/abs/2304.04065}{{\ttfamily
  arXiv:2304.04065 [astro-ph.CO]}}.

\bibitem{Choudhury:2023kdb}
S.~Choudhury, A.~Karde, S.~Panda, and M.~Sami, ``{Primordial non-Gaussianity
  from ultra slow-roll Galileon inflation},''
  \href{http://dx.doi.org/10.1088/1475-7516/2024/01/012}{{\em JCAP} {\bfseries
  01} (2024) 012}, \href{http://arxiv.org/abs/2306.12334}{{\ttfamily
  arXiv:2306.12334 [astro-ph.CO]}}.

\bibitem{Choudhury:2023hfm}
S.~Choudhury, A.~Karde, S.~Panda, and M.~Sami, ``{Scalar induced gravity waves
  from ultra slow-roll galileon inflation},''
  \href{http://dx.doi.org/10.1016/j.nuclphysb.2024.116678}{{\em Nucl. Phys. B}
  {\bfseries 1007} (2024) 116678},
  \href{http://arxiv.org/abs/2308.09273}{{\ttfamily arXiv:2308.09273
  [astro-ph.CO]}}.

\bibitem{Bhattacharya:2023ysp}
G.~Bhattacharya, S.~Choudhury, K.~Dey, S.~Ghosh, A.~Karde, and N.~S. Mishra,
  ``{Evading no-go for PBH formation and production of SIGWs using Multiple
  Sharp Transitions in EFT of single field inflation},''
  \href{http://dx.doi.org/10.1016/j.dark.2024.101602}{{\em Phys. Dark Univ.}
  {\bfseries 46} (2024) 101602},
  \href{http://arxiv.org/abs/2309.00973}{{\ttfamily arXiv:2309.00973
  [astro-ph.CO]}}.

\bibitem{Choudhury:2023fwk}
S.~Choudhury, K.~Dey, A.~Karde, S.~Panda, and M.~Sami, ``{Primordial
  non-Gaussianity as a saviour for PBH overproduction in SIGWs generated by
  pulsar timing arrays for Galileon inflation},''
  \href{http://dx.doi.org/10.1016/j.physletb.2024.138925}{{\em Phys. Lett. B}
  {\bfseries 856} (2024) 138925},
  \href{http://arxiv.org/abs/2310.11034}{{\ttfamily arXiv:2310.11034
  [astro-ph.CO]}}.

\bibitem{Choudhury:2023fjs}
S.~Choudhury, K.~Dey, and A.~Karde, ``{Untangling PBH overproduction in
  $w$-SIGWs generated by Pulsar Timing Arrays for MST-EFT of single field
  inflation},'' \href{http://arxiv.org/abs/2311.15065}{{\ttfamily
  arXiv:2311.15065 [astro-ph.CO]}}.

\bibitem{Choudhury:2024one}
S.~Choudhury, A.~Karde, S.~Panda, and M.~Sami, ``{Realisation of the ultra-slow
  roll phase in Galileon inflation and PBH overproduction},''
  \href{http://dx.doi.org/10.1088/1475-7516/2024/07/034}{{\em JCAP} {\bfseries
  07} (2024) 034}, \href{http://arxiv.org/abs/2401.10925}{{\ttfamily
  arXiv:2401.10925 [astro-ph.CO]}}.

\bibitem{Firouzjahi:2023ahg}
H.~Firouzjahi and A.~Riotto, ``{Primordial Black Holes and loops in
  single-field inflation},''
  \href{http://dx.doi.org/10.1088/1475-7516/2024/02/021}{{\em JCAP} {\bfseries
  02} (2024) 021}, \href{http://arxiv.org/abs/2304.07801}{{\ttfamily
  arXiv:2304.07801 [astro-ph.CO]}}.

\bibitem{Firouzjahi:2023aum}
H.~Firouzjahi, ``{One-loop corrections in power spectrum in single field
  inflation},'' \href{http://dx.doi.org/10.1088/1475-7516/2023/10/006}{{\em
  JCAP} {\bfseries 10} (2023) 006},
  \href{http://arxiv.org/abs/2303.12025}{{\ttfamily arXiv:2303.12025
  [astro-ph.CO]}}.

\bibitem{Riotto:2024ayo}
A.~Riotto and J.~Silk, ``{The Future of Primordial Black Holes: Open Questions
  and Roadmap},'' \href{http://arxiv.org/abs/2403.02907}{{\ttfamily
  arXiv:2403.02907 [astro-ph.CO]}}.

\bibitem{Choudhury:2024dei}
S.~Choudhury, A.~Karde, S.~Panda, and S.~SenGupta,
  ``{Regularized-Renormalized-Resummed loop corrected power spectrum of
  non-singular bounce with Primordial Black Hole formation},''
  \href{http://arxiv.org/abs/2405.06882}{{\ttfamily arXiv:2405.06882
  [astro-ph.CO]}}.

\bibitem{Choudhury:2024dzw}
S.~Choudhury, S.~Ganguly, S.~Panda, S.~SenGupta, and P.~Tiwari, ``{Obviating
  PBH overproduction for SIGWs generated by pulsar timing arrays in loop
  corrected EFT of bounce},''
  \href{http://dx.doi.org/10.1088/1475-7516/2024/09/013}{{\em JCAP} {\bfseries
  09} (2024) 013}, \href{http://arxiv.org/abs/2407.18976}{{\ttfamily
  arXiv:2407.18976 [astro-ph.CO]}}.

\bibitem{Choudhury:2024aji}
S.~Choudhury and M.~Sami, ``{Large fluctuations and Primordial Black Holes},''
  \href{http://arxiv.org/abs/2407.17006}{{\ttfamily arXiv:2407.17006 [gr-qc]}}.

\bibitem{Choudhury:2024kjj}
S.~Choudhury, K.~Dey, S.~Ganguly, A.~Karde, S.~K. Singh, and P.~Tiwari,
  ``{Negative non-Gaussianity as a salvager for PBHs with PTAs in bounce},''
  \href{http://arxiv.org/abs/2409.18983}{{\ttfamily arXiv:2409.18983
  [astro-ph.CO]}}.

\bibitem{Choudhury:2024ybk}
S.~Choudhury, ``{Large fluctuations in the sky},''
  \href{http://dx.doi.org/10.1142/S0218271824410074}{{\em Int. J. Mod. Phys. D}
  {\bfseries 33} no.~15, (2024) 2441007},
  \href{http://arxiv.org/abs/2403.07343}{{\ttfamily arXiv:2403.07343
  [astro-ph.CO]}}.

\bibitem{Gorbenko:2019rza}
V.~Gorbenko and L.~Senatore, ``{$\lambda \phi^4$ in dS},''
  \href{http://arxiv.org/abs/1911.00022}{{\ttfamily arXiv:1911.00022
  [hep-th]}}.

\bibitem{Cohen:2021fzf}
T.~Cohen, D.~Green, A.~Premkumar, and A.~Ridgway, ``{Stochastic Inflation at
  NNLO},'' \href{http://dx.doi.org/10.1007/JHEP09(2021)159}{{\em JHEP}
  {\bfseries 09} (2021) 159}, \href{http://arxiv.org/abs/2106.09728}{{\ttfamily
  arXiv:2106.09728 [hep-th]}}.

\bibitem{Cohen:2022clv}
T.~Cohen, D.~Green, and A.~Premkumar, ``{Large deviations in the early
  Universe},'' \href{http://dx.doi.org/10.1103/PhysRevD.107.083501}{{\em Phys.
  Rev. D} {\bfseries 107} no.~8, (2023) 083501},
  \href{http://arxiv.org/abs/2212.02535}{{\ttfamily arXiv:2212.02535
  [hep-th]}}.

\bibitem{Green:2022ovz}
D.~Green, ``{EFT for de Sitter Space},''
  \href{http://arxiv.org/abs/2210.05820}{{\ttfamily arXiv:2210.05820
  [hep-th]}}.

\bibitem{Cohen:2021jbo}
T.~Cohen, D.~Green, and A.~Premkumar, ``{A tail of eternal inflation},''
  \href{http://dx.doi.org/10.21468/SciPostPhys.14.5.109}{{\em SciPost Phys.}
  {\bfseries 14} no.~5, (2023) 109},
  \href{http://arxiv.org/abs/2111.09332}{{\ttfamily arXiv:2111.09332
  [hep-th]}}.

\bibitem{Cohen:2020php}
T.~Cohen and D.~Green, ``{Soft de Sitter Effective Theory},''
  \href{http://dx.doi.org/10.1007/JHEP12(2020)041}{{\em JHEP} {\bfseries 12}
  (2020) 041}, \href{http://arxiv.org/abs/2007.03693}{{\ttfamily
  arXiv:2007.03693 [hep-th]}}.

\bibitem{Starobinsky:1986fx}
A.~A. Starobinsky, ``{STOCHASTIC DE SITTER (INFLATIONARY) STAGE IN THE EARLY
  UNIVERSE},'' \href{http://dx.doi.org/10.1007/3-540-16452-9_6}{{\em Lect.
  Notes Phys.} {\bfseries 246} (1986) 107--126}.

\bibitem{Vennin:2024yzl}
V.~Vennin and D.~Wands, ``{Quantum diffusion and large primordial perturbations
  from inflation},'' \href{http://arxiv.org/abs/2402.12672}{{\ttfamily
  arXiv:2402.12672 [astro-ph.CO]}}.

\bibitem{Animali:2024jiz}
C.~Animali and V.~Vennin, ``{Clustering of primordial black holes from quantum
  diffusion during inflation},''
  \href{http://dx.doi.org/10.1088/1475-7516/2024/08/026}{{\em JCAP} {\bfseries
  08} (2024) 026}, \href{http://arxiv.org/abs/2402.08642}{{\ttfamily
  arXiv:2402.08642 [astro-ph.CO]}}.

\bibitem{LISACosmologyWorkingGroup:2023njw}
{\bfseries LISA Cosmology Working Group} Collaboration, E.~Bagui {\em et~al.},
  ``{Primordial black holes and their gravitational-wave signatures},''
  \href{http://arxiv.org/abs/2310.19857}{{\ttfamily arXiv:2310.19857
  [astro-ph.CO]}}.

\bibitem{Animali:2022otk}
C.~Animali and V.~Vennin, ``{Primordial black holes from stochastic
  tunnelling},'' \href{http://dx.doi.org/10.1088/1475-7516/2023/02/043}{{\em
  JCAP} {\bfseries 02} (2023) 043},
  \href{http://arxiv.org/abs/2210.03812}{{\ttfamily arXiv:2210.03812
  [astro-ph.CO]}}.

\bibitem{Ezquiaga:2022qpw}
J.~M. Ezquiaga, J.~Garc\'\i{}a-Bellido, and V.~Vennin, ``{Massive Galaxy
  Clusters Like El Gordo Hint at Primordial Quantum Diffusion},''
  \href{http://dx.doi.org/10.1103/PhysRevLett.130.121003}{{\em Phys. Rev.
  Lett.} {\bfseries 130} no.~12, (2023) 121003},
  \href{http://arxiv.org/abs/2207.06317}{{\ttfamily arXiv:2207.06317
  [astro-ph.CO]}}.

\bibitem{Jackson:2022unc}
J.~H.~P. Jackson, H.~Assadullahi, K.~Koyama, V.~Vennin, and D.~Wands,
  ``{Numerical simulations of stochastic inflation using importance
  sampling},'' \href{http://dx.doi.org/10.1088/1475-7516/2022/10/067}{{\em
  JCAP} {\bfseries 10} (2022) 067},
  \href{http://arxiv.org/abs/2206.11234}{{\ttfamily arXiv:2206.11234
  [astro-ph.CO]}}.

\bibitem{Tada:2021zzj}
Y.~Tada and V.~Vennin, ``{Statistics of coarse-grained cosmological fields in
  stochastic inflation},''
  \href{http://dx.doi.org/10.1088/1475-7516/2022/02/021}{{\em JCAP} {\bfseries
  02} no.~02, (2022) 021}, \href{http://arxiv.org/abs/2111.15280}{{\ttfamily
  arXiv:2111.15280 [astro-ph.CO]}}.

\bibitem{Pattison:2021oen}
C.~Pattison, V.~Vennin, D.~Wands, and H.~Assadullahi, ``{Ultra-slow-roll
  inflation with quantum diffusion},''
  \href{http://dx.doi.org/10.1088/1475-7516/2021/04/080}{{\em JCAP} {\bfseries
  04} (2021) 080}, \href{http://arxiv.org/abs/2101.05741}{{\ttfamily
  arXiv:2101.05741 [astro-ph.CO]}}.

\bibitem{Ando:2020fjm}
K.~Ando and V.~Vennin, ``{Power spectrum in stochastic inflation},''
  \href{http://dx.doi.org/10.1088/1475-7516/2021/04/057}{{\em JCAP} {\bfseries
  04} (2021) 057}, \href{http://arxiv.org/abs/2012.02031}{{\ttfamily
  arXiv:2012.02031 [astro-ph.CO]}}.

\bibitem{Ezquiaga:2019ftu}
J.~M. Ezquiaga, J.~Garc\'\i{}a-Bellido, and V.~Vennin, ``{The exponential tail
  of inflationary fluctuations: consequences for primordial black holes},''
  \href{http://dx.doi.org/10.1088/1475-7516/2020/03/029}{{\em JCAP} {\bfseries
  03} (2020) 029}, \href{http://arxiv.org/abs/1912.05399}{{\ttfamily
  arXiv:1912.05399 [astro-ph.CO]}}.

\bibitem{Pattison:2019hef}
C.~Pattison, V.~Vennin, H.~Assadullahi, and D.~Wands, ``{Stochastic inflation
  beyond slow roll},''
  \href{http://dx.doi.org/10.1088/1475-7516/2019/07/031}{{\em JCAP} {\bfseries
  07} (2019) 031}, \href{http://arxiv.org/abs/1905.06300}{{\ttfamily
  arXiv:1905.06300 [astro-ph.CO]}}.

\bibitem{Noorbala:2018zlv}
M.~Noorbala, V.~Vennin, H.~Assadullahi, H.~Firouzjahi, and D.~Wands,
  ``{Tunneling in Stochastic Inflation},''
  \href{http://dx.doi.org/10.1088/1475-7516/2018/09/032}{{\em JCAP} {\bfseries
  09} (2018) 032}, \href{http://arxiv.org/abs/1806.09634}{{\ttfamily
  arXiv:1806.09634 [hep-th]}}.

\bibitem{Pattison:2017mbe}
C.~Pattison, V.~Vennin, H.~Assadullahi, and D.~Wands, ``{Quantum diffusion
  during inflation and primordial black holes},''
  \href{http://dx.doi.org/10.1088/1475-7516/2017/10/046}{{\em JCAP} {\bfseries
  10} (2017) 046}, \href{http://arxiv.org/abs/1707.00537}{{\ttfamily
  arXiv:1707.00537 [hep-th]}}.

\bibitem{Grain:2017dqa}
J.~Grain and V.~Vennin, ``{Stochastic inflation in phase space: Is slow roll a
  stochastic attractor?},''
  \href{http://dx.doi.org/10.1088/1475-7516/2017/05/045}{{\em JCAP} {\bfseries
  05} (2017) 045}, \href{http://arxiv.org/abs/1703.00447}{{\ttfamily
  arXiv:1703.00447 [gr-qc]}}.

\bibitem{Hardwick:2017fjo}
R.~J. Hardwick, V.~Vennin, C.~T. Byrnes, J.~Torrado, and D.~Wands, ``{The
  stochastic spectator},''
  \href{http://dx.doi.org/10.1088/1475-7516/2017/10/018}{{\em JCAP} {\bfseries
  10} (2017) 018}, \href{http://arxiv.org/abs/1701.06473}{{\ttfamily
  arXiv:1701.06473 [astro-ph.CO]}}.

\bibitem{Mishra:2023lhe}
S.~S. Mishra, E.~J. Copeland, and A.~M. Green, ``{Primordial black holes and
  stochastic inflation beyond slow roll. Part I. Noise matrix elements},''
  \href{http://dx.doi.org/10.1088/1475-7516/2023/09/005}{{\em JCAP} {\bfseries
  09} (2023) 005}, \href{http://arxiv.org/abs/2303.17375}{{\ttfamily
  arXiv:2303.17375 [astro-ph.CO]}}.

\bibitem{Choudhury:2024jlz}
S.~Choudhury, A.~Karde, P.~Padiyar, and M.~Sami, ``{Primordial black holes from
  effective field theory of stochastic single field inflation at NNNLO},''
  \href{http://dx.doi.org/10.1140/epjc/s10052-024-13644-2}{{\em Eur. Phys. J.
  C} {\bfseries 85} no.~1, (2025) 21},
  \href{http://arxiv.org/abs/2403.13484}{{\ttfamily arXiv:2403.13484
  [astro-ph.CO]}}.

\bibitem{Weinberg:2008hq}
S.~Weinberg, ``{Effective Field Theory for Inflation},''
  \href{http://dx.doi.org/10.1103/PhysRevD.77.123541}{{\em Phys. Rev. D}
  {\bfseries 77} (2008) 123541},
  \href{http://arxiv.org/abs/0804.4291}{{\ttfamily arXiv:0804.4291 [hep-th]}}.

\bibitem{Cheung:2007st}
C.~Cheung, P.~Creminelli, A.~L. Fitzpatrick, J.~Kaplan, and L.~Senatore, ``{The
  Effective Field Theory of Inflation},''
  \href{http://dx.doi.org/10.1088/1126-6708/2008/03/014}{{\em JHEP} {\bfseries
  03} (2008) 014}, \href{http://arxiv.org/abs/0709.0293}{{\ttfamily
  arXiv:0709.0293 [hep-th]}}.

\bibitem{Choudhury:2017glj}
S.~Choudhury, ``{CMB from EFT},''
  \href{http://dx.doi.org/10.3390/universe5060155}{{\em Universe} {\bfseries 5}
  no.~6, (2019) 155}, \href{http://arxiv.org/abs/1712.04766}{{\ttfamily
  arXiv:1712.04766 [hep-th]}}.

\bibitem{Delacretaz:2016nhw}
L.~V. Delacretaz, V.~Gorbenko, and L.~Senatore, ``{The Supersymmetric Effective
  Field Theory of Inflation},''
  \href{http://dx.doi.org/10.1007/JHEP03(2017)063}{{\em JHEP} {\bfseries 03}
  (2017) 063}, \href{http://arxiv.org/abs/1610.04227}{{\ttfamily
  arXiv:1610.04227 [hep-th]}}.

\bibitem{Naskar:2017ekm}
A.~Naskar, S.~Choudhury, A.~Banerjee, and S.~Pal, ``{EFT of Inflation:
  Reflections on CMB and Forecasts on LSS Surveys},''
  \href{http://arxiv.org/abs/1706.08051}{{\ttfamily arXiv:1706.08051
  [astro-ph.CO]}}.

\bibitem{Enqvist:2008kt}
K.~Enqvist, S.~Nurmi, D.~Podolsky, and G.~I. Rigopoulos, ``{On the divergences
  of inflationary superhorizon perturbations},''
  \href{http://dx.doi.org/10.1088/1475-7516/2008/04/025}{{\em JCAP} {\bfseries
  04} (2008) 025}, \href{http://arxiv.org/abs/0802.0395}{{\ttfamily
  arXiv:0802.0395 [astro-ph]}}.

\bibitem{Fujita:2013cna}
T.~Fujita, M.~Kawasaki, Y.~Tada, and T.~Takesako, ``{A new algorithm for
  calculating the curvature perturbations in stochastic inflation},''
  \href{http://dx.doi.org/10.1088/1475-7516/2013/12/036}{{\em JCAP} {\bfseries
  12} (2013) 036}, \href{http://arxiv.org/abs/1308.4754}{{\ttfamily
  arXiv:1308.4754 [astro-ph.CO]}}.

\bibitem{Fujita:2014tja}
T.~Fujita, M.~Kawasaki, and Y.~Tada, ``{Non-perturbative approach for curvature
  perturbations in stochastic $\delta N$ formalism},''
  \href{http://dx.doi.org/10.1088/1475-7516/2014/10/030}{{\em JCAP} {\bfseries
  10} (2014) 030}, \href{http://arxiv.org/abs/1405.2187}{{\ttfamily
  arXiv:1405.2187 [astro-ph.CO]}}.

\bibitem{Vennin:2015hra}
V.~Vennin and A.~A. Starobinsky, ``{Correlation Functions in Stochastic
  Inflation},'' \href{http://dx.doi.org/10.1140/epjc/s10052-015-3643-y}{{\em
  Eur. Phys. J. C} {\bfseries 75} (2015) 413},
  \href{http://arxiv.org/abs/1506.04732}{{\ttfamily arXiv:1506.04732
  [hep-th]}}.

\bibitem{Dias:2012qy}
M.~Dias, R.~H. Ribeiro, and D.~Seery, ``{The \ensuremath{\delta}N formula is
  the dynamical renormalization group},''
  \href{http://dx.doi.org/10.1088/1475-7516/2013/10/062}{{\em JCAP} {\bfseries
  10} (2013) 062}, \href{http://arxiv.org/abs/1210.7800}{{\ttfamily
  arXiv:1210.7800 [astro-ph.CO]}}.

\end{thebibliography}\endgroup
\bibliographystyle{utphys}


\end{document}